\documentclass[fleqn,usenatbib]{mnras}

\usepackage[T1]{fontenc}
\usepackage{ae,aecompl}

\usepackage{graphicx}	
\usepackage{amsmath}	
\usepackage{amssymb}	
\usepackage{CJK}

\newcommand\eg{{\it e.g.,} }

\title[ ]{Host-star and exoplanet compositions: a pilot study using a wide binary with a polluted white dwarf}

\author[]{Amy Bonsor$^{1}$\thanks{E-mail: abonsor@ast.cam.ac.uk}
Paula Jofr\'e$^{2}$, Oliver Shorttle$^{1,3}$,  Laura K  Rogers$^{1}$, \newauthor
Siyi Xu(许\CJKfamily{bsmi}偲\CJKfamily{gbsn}艺)$^{4}$ and Carl Melis$^5$\\
$^{1}$Institute of Astronomy, University of Cambridge, Madingley Road, Cambridge, CB3 0HA, UK \\
$^{2}$N\'{u}cleo de Astronom\'{i}a, Universidad Diego Portales, Ej\'{e}rcito 441, Santiago de Chile\\
$^{3}$Department of Earth Sciences, University of Cambridge, Cambridge CB2 3EQ, UK\\
$^{4}$Gemini Observatory/NSF's NOIRLab, 670 N. A’ohoku Place, Hilo, HI 96720, USA\\
$^{5}$Center for Astrophysics and Space Sciences, University of California, San Diego, CA 92093-0424, USA. }

\date{Accepted XXX. Received YYY; in original form ZZZ}

\pubyear{2020}

\begin{document}
\begin{CJK}{UTF8}{gbsn}
\label{firstpage}
\pagerange{\pageref{firstpage}--\pageref{lastpage}}
\maketitle

\begin{abstract}
Planets and stars ultimately form out of the collapse of the same cloud of gas. Whilst planets, and planetary bodies, readily loose volatiles, a common hypothesis is that they retain the same refractory composition as their host star. This is true within the Solar System. The refractory composition of chondritic meteorites, Earth and other rocky planetary bodies are consistent with solar, within the observational errors. This work aims to investigate whether this hypothesis holds for exoplanetary systems. If true, the internal structure of observed rocky exoplanets can be better constrained using their host star abundances. In this paper, we analyse the abundances of the K-dwarf, G200-40, and compare them to its polluted white dwarf companion, WD 1425+540. The white dwarf has accreted planetary material, most probably a Kuiper belt-like object, from an outer planetary system surviving the star's evolution to the white dwarf phase. Given that binary pairs are chemically homogeneous, we use the binary companion, G200-40, as a proxy for the composition of the progenitor to WD 1425+540. We show that the elemental abundances of the companion star and the planetary material accreted by WD 1425+540 are consistent with the hypothesis that planet and host-stars have the same true abundances, taking into account the observational errors.

\end{abstract}



\section{Introduction}

The processes occuring during planet formation have been key to our own planet's history, structure and its development of life, just as they have been key in determining the nature of planets around other stars. We lack a full understanding of how planet formation determines the composition of a planet. Whilst planets clearly form out of the same material as their host stars, a range of processes may occur during planet formation and evolution that alter compositions. Volatiles are easily lost, or are less readily incorporated, into planetary bodies at high temperatures. However, not all planetary bodies have experienced high temperatures. In fact, many comets in the Solar System retain clear signatures of their origin in interstellar gas \citep{Mumma2011}. 

The refractory content of planetary bodies will by definition not be influenced by the high temperatures experienced, even in the inner regions of planet-forming discs. In our Solar System, many planetary bodies match the refractory composition of the Sun. Modelling of the interior structure of Pluto and Charon from New Horizons provided no evidence that the interior had a composition different from solar \citep{McKinnon2017}. The refractory composition of chondritic meteorites and the Sun agree to within errors on the solar abundances \citep{Anders1982}. Indeed a chondritic reference model has proved a powerful tool for studying the bulk composition of Earth and the terrestrial planets \citep[\eg][]{Palme2003treatise}, although there are many subtle variations in elemental and isotopic ratios which point to the complex nature of planet formation \citep{Boyet2005,Burkhardt2011,Klaver2020}. For rocky exoplanets, a match in the refractory content of the planet and host star, can significantly improve our ability to determine the interior structure of an observed planet based on its density alone \citep{Dorn2015, Dorn2017a}. Given that for most rocky exoplanets, mass and radii measurements will remain the best probe of their interior structure, this assumption can make a key difference in our ability to characterise detected planets.

Like stars and planets, pairs of binary stars should form out of the collapse of the same cloud of interstellar gas. Thus, even if early in their lifetimes, in the centre of dense stellar clusters, scattering and exchange of binary pairs is common \citep{MalmbergDavies09}, binary pairs should have very similar compositions, especially true for wide binaries that have not interacted. Recent works confirm this trend, \eg \citet{andrews2019, Hawkins2020} show that wide binary stars are more chemically homogeneous than random pairs of stars in the Galactic disk. These findings are consistent with recent complementary work on chemical abundances of open clusters \citep{bovy2016, casamiquela2020}, showing that member stars are more chemically homogeneous than field stars.

Here, we use this principle of chemical homogeneity in stars born together to present a novel way to test whether the refractory abundances of host stars match their planets. We compare the composition of exoplanetary bodies and their wide binary companion. If the refractory composition of the planetary bodies matches those of their wide binary companion, this indicates that the refractory composition of the planetary bodies and their host star are likely to be the same, given the observed chemical homogeneity of wide binary pairs. We present observations and abundance analysis for the K-dwarf, G200-40, and compare with abundances derived for the planetary material in the atmosphere of its wide binary companion, the polluted white dwarf, WD 1425+540.  

More than a thousand white dwarfs, known as {\it polluted}, have metals from planetary material in their atmospheres \citep{Coutu2019}. Given that white dwarfs should have clean hydrogen or helium atmospheres, with metals heavier than helium sinking out of sight on timescales (days to millions of years) much shorter than the white dwarf cooling time  \citep{Koester09}, the presence of metals in the atmospheres of 30-50\% of white dwarfs \citep{ZK10,Koester2014} suggests the recent accretion of planetary bodies. Outer planetary systems should survive intact to the white dwarf phase and planetary bodies scattered inwards can be tidally disrupted and accreted by the white dwarf \citep{Veras_review}. Observations of polluted white dwarfs can tell us about the range of compositions present in exoplanetary bodies, as well as the geology of rocky exoplanets and processes that alter planetary compositions \citep{JuraYoung2014}. Evidence exists for white dwarfs that have accreted icy planetary bodies, rocky planetary bodies and even planetary bodies very rich in highly refractory material similar to calcium aluminium rich inclusions \citep[\eg][]{Farihi2011_water, Zuckerman2011, Harrison2018}.

WD 1425+540 is thought to have accreted an icy, volatile-rich body, whose best analog in the Solar System is comet Halley \citep{Xu2017}. Abundances for the elements C, N, O, Mg, Si, S, Ca, Fe, and Ni are derived from its spectra \citep{Xu2017}. WD 1425+540 has a helium dominated atmosphere and an effective temperature of 14,490K \citep{Bergeron2011}. In this work, we present observations of the K-dwarf, G200-40, whose abundances are to be compared to its wide binary companion, the highly polluted white dwarf, WD 1425+540. The two stars are separated by $\sim~40$ arcsec (projected separation 2240 au) in the sky \citep{Wegner1981}. Gaia photometry confirms that they are a common proper motion pair. In this paper, we compare the abundances of refractory species in G200-40 to those determined for WD 1425+540, with the aim of probing whether refractory compositions of exoplanetary bodies really do match those of their host-star. We present the observations and derive abundances for the companion, G200-40 in \S\ref{sec:observations}. This is followed by a comparison with the abundances derived previously for WD 1425+540 in \S\ref{sec:results} and a discussion of the implications for our understanding of planet formation and future observations in \S\ref{sec:discussion}.

\section{Observations and Abundance determination for G200-40 and 61 Cyg B}
\label{sec:observations}

Optical spectroscopic observations were conducted at Lick Observatory with the 
2.4\,m Automated Planet Finder telescope (APF; \citealt{vogt14}).
APF feeds the Levy Spectrograph, a stabilized instrument with
a minimum of moveable parts that covers most of the optical wavelength range. 
A 1.0$'$$'$ slit was used resulting
in resolving powers of $\gtrsim$100,000.
Light is recorded on an E2V CD42-90 back-illuminated CCD which in practice
produces strong fringing patterns for wavelengths longward of 7000\,\AA ; as a result,
spectral orders beyond 7000\,\AA\ are not used for abundance determinations.

G200-40 
was observed on UT 2017 August 11. Three exposures of 3000~seconds
each were obtained at the beginning of the night resulting in a signal-to-noise
ratio of $\approx$60 at a wavelength of 6000\,\AA . At the end of the night, exposures
of similar K-dwarf, 61\,Cyg~B were obtained with the integration time limited by the photon
integrator (set to obtain signal-to-noise $>$100 in most orders).

Data reduction is performed with standard {\sf IRAF} echelle tasks.
Two-dimensional spectral image frames are bias subtracted, flat-fielded, extracted, 
and finally wavelength calibrated with ThAr arclamp spectra.

\begin{table*}    \centering
    \begin{tabular}{c|c|c|c}
    \hline
 
Name & WD 1425+540 & G200-40 & 61 Cyg B\\ \hline
Source ID & 1608497864040134016 & 1608497623521965184 & 1872046574983497216 \\
$\alpha$ (deg) & 216.89821430610 & 216.90195396444. & 316.75293147546  \\
$\delta$ (deg) & +53.80838190929 & +53.79756047297. & +38.75563452356	  \\

     \end{tabular}
    \caption{\textit{Gaia} DR2 astrometry for the target K-dwarf, G200-40, its white dwarf companion, WD 1425+540 and the comparison K-dwarf, 61 Cyg B. $\alpha$ and $\delta$ are barycentric and in ICRS at \textit{Gaia} Epoch 2015.5. }
    \label{tab:astrometry}
\end{table*}

The spectral analysis was done using the {\tt iSpec} framework \citep{ispec}, which performs synthesis  on-the-fly on pre-defined spectral regions until a good fit to the observed spectrum is found. {\tt iSpec} is a wrapper code that has the option to call several state-of-the-art radiative transfer codes and model atmospheres to perform the syntheses. Here, we employed the code TURBOSPECTRUM \citep{turbospectrum} and the MARCS models \citep{marcs}, which consider local thermodynamical equilibrium and one-dimensional atmospheric layers. We considered the atomic data from VALD as well as the line list developed for the Gaia-ESO survey \citep{Heiter2020}.

A crucial part of the analysis is to find the set of suitable regions in the spectrum for the determination of atmospheric parameters and then for the abundances \citep[\eg][]{jofre19}. These regions need to show dependency on temperature, surface gravity, metallicity as well as the elements measured in the white dwarf. Therefore, we used the spectra of the K-dwarf Gaia benchmark star 
61 Cyg B \citep{jofre18}, as well as the Sun, to help us to identify the spectral regions for our analysis. These stars are widely used as reference pillars for spectroscopic analysis of automatic pipelines in stellar surveys.  
The spectral regions included the wings of strong lines, in addition to several lines of iron-peak and alpha elements.

 For G200-40, we determined the following atmospheric parameters.  We obtained an effective temperature $T_{\rm eff} = 4036 \pm  66$K, a surface gravity  $\log g = 4.34  \pm 0.48$ dex, a total metal content\footnote{For solar-scaled abundances, the metal content [M/H] and the iron content [Fe/H] is the same, but for metal-poor alpha-enhanced stars, [M/H] is slightly higher than [Fe/H]. When deriving metallicities from alpha-capture elements, it is important to be aware of this potential difference if the abundances are not solar-scaled. }  $[{\rm M/H}] = -0.63
\pm 0.18$, and microturbulence parameter $v_t = 0.75 \pm 0.48$ km\,s$^{-1}$.  We further determined a stellar mass of $0.561 \pm  0.009M_\odot$ and radius of $0.526 \pm  0.007R_\odot$. The uncertainties are internal only, and their large values reflect the difficulties in obtaining a good fit for this spectrum.  The cool temperatures imply the spectrum is crowded with molecules, making it difficult to normalise the continuum and find regions free of molecules \citep[see \eg][]{lebzelter12, jofre15}. Although we include molecules in our line list, the list is still incomplete for such cool stars \citep{masseron14}.

To validate our results, we used the same procedure to analyse 61 Cyg B finding parameters of $T_{\rm eff} = 4064 \pm  50$K, $\log g = 4.33  \pm 0.2$ dex, $[M/H] = -0.61 \pm 0.2$ and $v_t = 0.64 \pm 0.1$ km\,s$^{-1}$, where uncertainties are estimated, but the final values agree within the uncertainties with those reported by \cite{jofre18}. We note, however, that our metal content is lower. This can be explained because the value determined by \cite{jofre18} considered a line list without molecules for the iron abundance determination. That work  reported a final result which was a combination of several different methods. One such methods included molecules, and that result yields a metallicity with excellent agreement with our result \citep[see][for details about this method and the procedure in]{jofre14}. We, therefore, conclude that although our parameters have high uncertainties, our spectral analysis yields accurate results.

To derive the abundances, we searched by eye all possible atomic spectral lines in the visual Atlas of Acturus \citep{atlas} that are visible in our spectrum. Then, we selected only the lines that had a fit with a conservative uncertainty below 1.5 dex and a root-mean-square of the difference between synthesis and observation for a given line below 0.1. We could not derive abundances of S, C and O  from atomic lines, their uncertainties were too large or the lines were too blended. Therefore, we performed syntheses for a star with stellar parameters obtained before but varying  O, C, N, Mg and Si to identify possible molecular regions that could serve as alternative. We were still unable to derive abundances for N and S. Our final values are shown in Table~\ref{tab:G200-40} for G200-40 and 61 Cyg  B, indicating the number of lines or regions for the respective abundances. The final abundances are considered to be the median with the standard deviation of the distribution as the uncertainty.  

For C and Mg we can compare our results obtained from atoms and molecules. The value with an asterisk in Table~\ref{tab:G200-40} is the adopted one. For carbon, the only atomic line (5052.19 \AA) was quite blended and yielded an uncertain result. Furthermore, the resulting value is  0.1 dex higher than the result obtained with molecules and too high for our expectations of 61 Cyg B, given its metallicity.  Therefore, we concluded that the C abundance from that atomic line was less accurate than the result from molecular features. 

For the case of Mg, however, we adopted the atomic value, even if its uncertainties might seem larger. The Mg lines adopted here (5528.43 \AA\  and 5711.09 \AA) are widely used for Mg determination for galactic stellar populations. On the other hand, very little has been reported in terms of abundance determination from Mg molecular features. Since the values do not agree between the atomic and molecular Mg obtained for G200-40, and the atomic Mg value obtained for 61 Cyg B agrees well with our reference taken from \cite{jofre15}, we adopted the conservative value, namely the atomic one.



\begin{table}
\centering
    \begin{tabular}{c|c|c|c|c}
    \hline
       Element & Atoms & No. of  & Molecules & No. of \\
       & & lines & & lines \\ \hline
       \multicolumn{3}{l}{G200-40 }
       \\ \hline
    $[{\rm C/H}]$  & $-0.12 \pm 0.14$ &1& $-0.27^* \pm 0.07$ & 6 \\
        $[{\rm O/H}]$ & & & $-0.15 \pm 0.07$ &13 \\
     $[{\rm Ca/H}]$ & $-0.37 \pm 0.04$& 22&&\\
    $[{\rm Fe/H}]$ & $-0.49  \pm  0.03$ & 101 &\\
    $[{\rm Mg/H}]$ & $-0.64^* \pm 0.14 $ &2  &  $-0.80 \pm 0.05$ &14 \\ 
    $[{\rm Ni/H}]$ & $-0.42 \pm 0.09$ & 13 \\
$[{\rm Si/H}]$ & &&$-0.45 \pm 0.18$ &2 \\ \hline
\multicolumn{3}{l}{61 Cyg B }\\ \hline
    $[{\rm C/H}]$  & $-0.18 \pm 0.12$ &1& $-0.24^* \pm 0.05$ & 5 \\
        $[{\rm O/H}]$ & & & $-0.59 \pm 0.09$ &18 \\
     $[{\rm Ca/H}]$ & $-0.52 \pm 0.03$& 22&&\\
    $[{\rm Fe/H}]$ & $-0.66  \pm  0.03$ & 101 &\\
    $[{\rm Mg/H}]$ & $-0.68^* \pm 0.24 $ &2  &  $-0.68 \pm 0.05$ &15 \\ 
    $[{\rm Ni/H}]$ & $-0.60 \pm 0.20$ & 13 \\
$[{\rm Si/H}]$ & &&$-0.48 \pm 0.32$ &2 \\
    \end{tabular}
    \caption{Abundances for G200-40 and 61 Cyg B. * indicates the adopted abundances. }
    \label{tab:G200-40}
\end{table}

\begin{table}
    \centering
    \begin{tabular}{c|c}
       \hline  Element & $\log  n({\rm Z})/n({\rm He})$\\ \hline
H &$-4.20\pm0.10$ \\
C  &$-7.29\pm0.17$\\
N  &$-8.09\pm0.10$\\
O &$ -6.62\pm0.23$\\
Mg  &$-8.16\pm0.20$\\ 
Si &$-8.03\pm0.31$\\
S  &$-8.36\pm0.11$\\ 
Ca  &$-9.26\pm0.10$\\
Fe  &$-8.15\pm0.14$\\ 
Ni  &$-9.67\pm0.20 $\\
\hline
    \end{tabular}
    \caption{The log number abundances relative to He derived for the planetary material accreted by WD 1425+540 taken from \citet{Xu2017}, assuming that $\log n({\rm H})/n({\rm He})=-4.2$, as derived from Balmer lines. }
    \label{tab:wd1425}
\end{table}

\begin{figure*}
\includegraphics[width=0.8\textwidth]{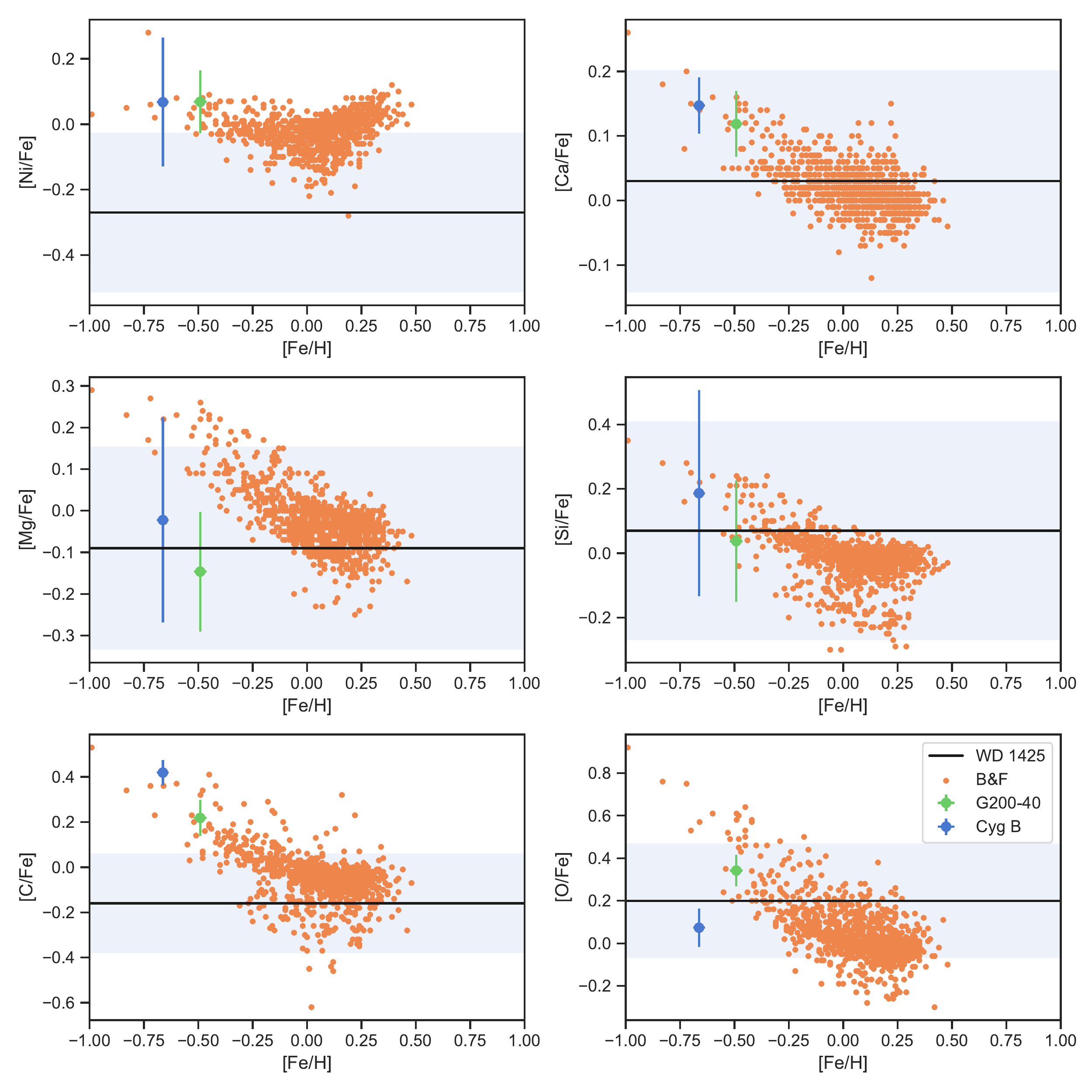}
\caption{The abundances relative to solar determined for G200-40, 61 Cyg B as a function of their metallicity (Table~\ref{tab:G200-40}), with $1-\sigma$ error bars. The abundances for WD 1425 +540 are shown as a black line and blue shaded region ($1-\sigma$ error bars), as the [Fe/H] cannot be compared. The abundances have not been adjusted for relative sinking of different species, as we assume that the planetary material is accreting in build-up phase.  Stars from \citet{Brewer2016} are included in orange for comparison, noting that typical errors are quoted as being between 0.01 and 0.04 dex. G200-40 has similar abundances to other metal poor stars.}
\label{fig:star_h}
\end{figure*}

\begin{figure}
\includegraphics[width=0.45\textwidth]{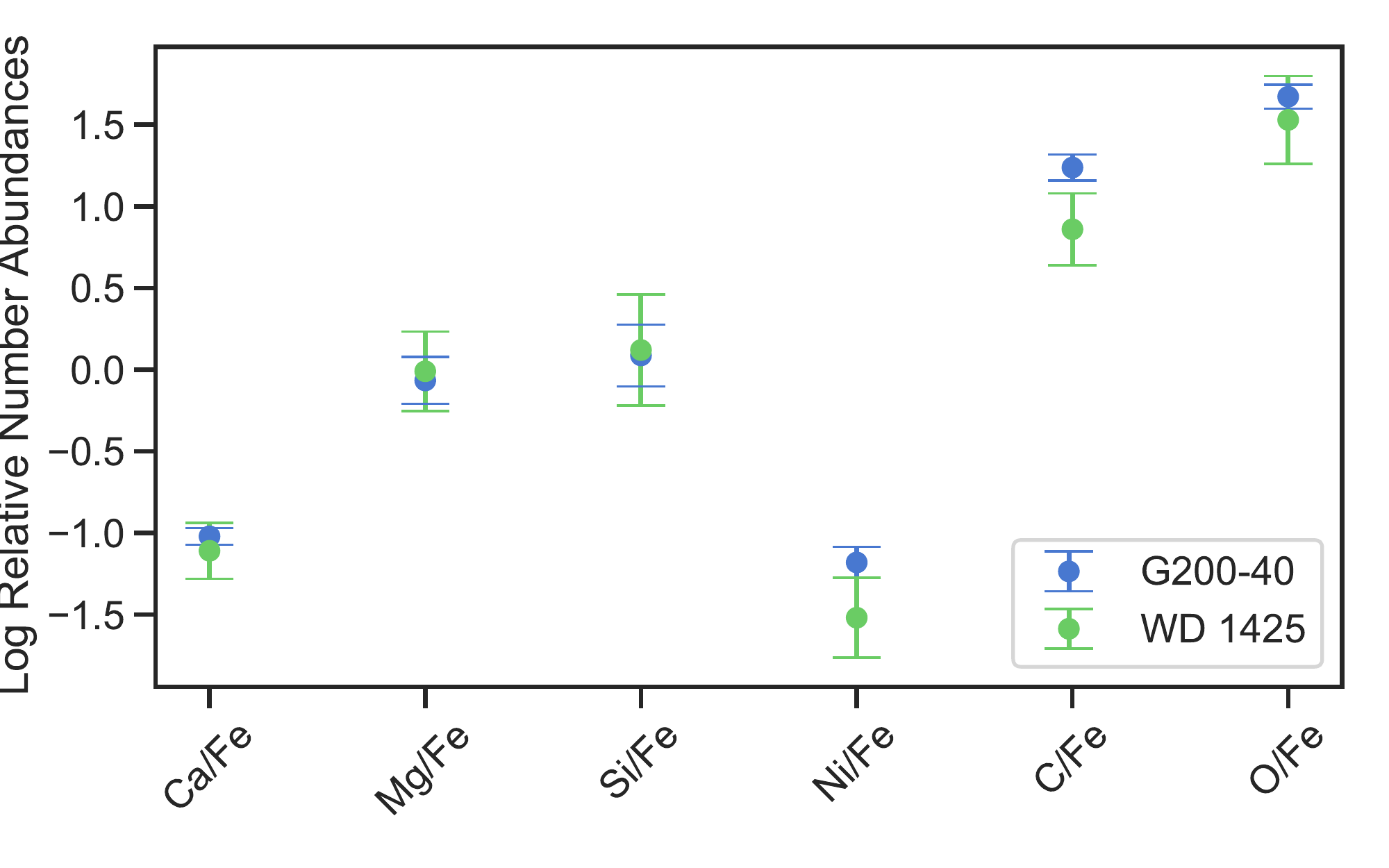}
\caption{The log number abundances relative to iron for WD 1425+540 and G200-40, alongside $1-\sigma$ error bars. The white dwarf abundances have not been adjusted for the relative sinking, as we assume that the planetary material is accreting in build-up phase. All abundances, except C/Fe and Ni/Fe are consistent to within $1-\sigma$. }
\label{fig:comparison}
\end{figure}

\section{Results}
\label{sec:results}
The final abundances for 61 Cyg B and G200-40 are presented in Table.~\ref{tab:G200-40}. As both 
61 Cyg B and G200-40 have very similar stellar parameters, and the same techniques were used to derive their abundances, we can compare their relative abundances to higher accuracy than to stars with very different properties \citep[see \eg][]{jofre15, jofre19, casamiquela2020}. Fig.~\ref{fig:star_h} shows the abundances relative to solar plotted against their stellar metallicity [Fe/H]. Both G200-40 and 61 Cyg B are relatively metal poor ($[{\rm Fe/H}]=-0.49$ or $-0.66$ or $[{\rm M/H}] = -0.61,-0.63$) and their abundances are fairly consistent with those of other similar metal-poor stars in \citet{Brewer2016}, shown in orange on Fig.~\ref{fig:star_h} for comparison. We note here that the Mg for G200-40 and O for 61 Cyg B do not sit as comfortably with the \cite{Brewer2016} sample as other elements, potentially due to uncertainties in our abundance determination. \cite{jofre15} discuss in detail the challenges involved in deriving abundances for the cool dwarfs analysed here in comparison with warmer solar-type stars such as those analysed by \cite{Brewer2016}. Issues of note include that the two Mg lines detected here were very blended and that our spectra do not reach the 777 nm range required to determine O abundance from the O triplet and as a result molecular features must be used instead.  

The abundances for WD 1425+540 are listed in Table~\ref{tab:wd1425}. We note here that we use the abundances labelled Model I in \cite{Xu2017}, derived using log n(H)/n(He) = -4.2, as recent studies show that the hydrogen abundance derived using Balmer lines is more reliable \citep{Gaensicke2018, Allard2020}. An additional complication regards the sinking of elemental species out of the white dwarf atmosphere. If the accretion started recently (compared to typical sinking timescales on the order of 1\,Myr), the system will be in build-up phase and the abundances will reflect those of the accreted body. On the other hand, if accretion has reached a steady-state between diffusion and accretion, the abundances of the accreted planetary material will have been adjusted by its sinking timescale.

\cite{Harrison2018} model in detail the composition of the accreted material and find that the abundances are best explained if accretion is in the build-up phase. If WD 1425+540 were accreting in steady-state, Ca/Fe in the accreted planetary body would be significantly higher than Ca/Fe in nearby stars. Ca/Fe can be enhanced in planetary bodies due to processing at high temperatures and the depletion of iron-rich minerals.  However, this would be at odds with the high volatile abundance (C, N, O and S) of the accreted planetary body. Ca/Fe is also enriched in the mantles of differentiated planetary bodies. However, in this case, Mg/Fe and other lithophile/sidereophile abundance ratios would be enhanced, which is not the case. For the rest of this paper, therefore, we assume that WD 1425+540 is accreting in the build-up phase.

\begin{figure*}
\includegraphics[width=0.98\textwidth]{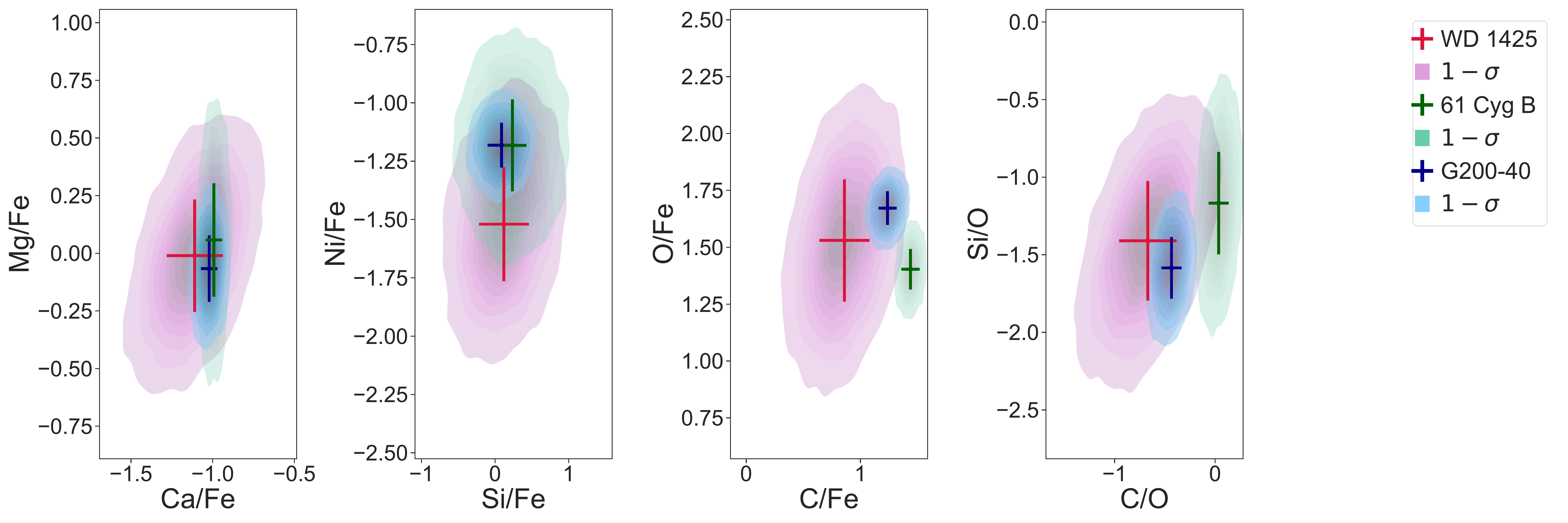}
\caption{The relative log number abundance ratios for G200-40 (blue), its white dwarf companion, WD 1145+540 (red), and a similar star, 61 Cyg B (green), listed in Table~\ref{tab:G200-40}. The pink, blue and green shaded regions indicate the correlated error structure. Seven observed elements leads to six independent ratios to Fe, which are shown on the three left-most panels. In addition to which, the right-hand panel shows, Si/O against C/O. }
\label{fig:ratios}
\end{figure*}

In order to compare the abundances observed for the planetary material in the atmosphere of the white dwarf WD 1425+540 with those of its companion, we consider metals only (Z$>$He). The abundances for WD 1425+540 are shown as black horizontal lines, with $1\sigma$ error regions shaded in blue, on Fig.~\ref{fig:star_h}. Fig.~\ref{fig:comparison} shows the log number abundances relative to iron for all 6 species observed in both WD 1425+540 and G200-40. All number abundances are consistent within $1\sigma$, apart from C/Fe and Ni/Fe, which are consistent within $2\sigma$. Fig.~\ref{fig:star_h} highlights how the comparison is limited 
by the uncertainties on both the white dwarf and K-dwarf abundances. Often the $1\sigma$ uncertainties on the white dwarf abundances span many of the typical stellar abundances from \cite{Brewer2016}. Notably the white dwarf abundances sit centrally  compared to the stars in \cite{Brewer2016}, apart from Ni/Fe, a ratio that can be altered due to planetary processing, for example segregation into iron melt.  Fig.~\ref{fig:ratios} shows the abundance ratios, alongside their correlated error structures. There is significant overlap in the abundances for WD 1425+540 and G200-40, whilst between WD 1425+540 and 61 CyB the overlap is more limited, particularly in such ratios as C/Fe, O/Fe, C/O and Si/O.

 We next quantify how likely it is that the true abundances of the star G200-40 match those of the material polluting its white dwarf companion, given the observational errors, and across multiple elemental ratios.  For this analysis we create a set of synthetic abundances, based on the observational errors and an assumption that the true abundances match and ask how often these synthetic observed abundances lie closer together than the actual observed abundances.

 In order to make this comparison, we must first transform the data via isometric log ratio, such that we take into account all elemental ratios equally, and we consider the full covariance error matrix, similar to those shown in Fig.~\ref{fig:ratios}. We find that 70\% of $10^4$ randomly sampled synthetic abundances are closer together (in terms of their Mahalanobis distance) than the observed abundances. Thus, if the material accreted by the white dwarf and its companion truly had the same elemental abundances, then given the error on the observations, 70\% of the time that they were observed, we would find abundances that are less similar than those reported here.

 In other words, this supports G200-40 having the same refractory element composition as the planetary material accreted by its white dwarf companion. The same exercise applied to 61 Cyg B finds that $<25\%$ of the samples are as far, or further, apart than the real data. Thus, we can reject the hypothesis that the abundances observed for 61 Cyg B match WD 1425 +540 at a 75\% confidence level.

\section{Discussion} 
\label{sec:discussion}
In this work, we present a pilot study to investigate whether the composition of rocky exoplanets matches that of their host stars. We present abundance analysis for the K-dwarf, G200-40, and compare with the abundances of the planetary material found in the atmosphere of its white dwarf companion, WD 1425+540. Assuming that the abundances of G200-40 are a good proxy for those of the progenitor to WD 1425+540, this technique can be used to compare the abundances of rocky exoplanets and their host-stars. 

\subsection{White dwarfs as tracers of planetary compositions}
This work is based on the premise that the abundances measured in the atmospheres of white dwarfs provide good tracers of the composition of planetary bodies from an outer planetary system. Whilst our understanding of the accretion of planetary bodies by white dwarfs is good, many details are missing \citep{Farihi_review}. For the purpose of this work, we assume that abundances are not altered during accretion, although current models do not yet rule this possibility out. Whether multiple, or a single planetary body is seen in the white dwarf atmosphere simultaneously does not change our conclusions, as the sum of the refractory abundances of many bodies should still match that of the host-star. We noted earlier in the text, however, that the phase of accretion is crucial. We hypothesise that WD 1425+540 started accreting planetary material recently (in the last 1\,Myr) and is accreting in build-up phase, such that the accreted abundances match those of known planetary bodies \citep{Harrison2018}. However, if the system could be shown to be accreting in steady-state, the high Ca/Fe and Ni/Fe ratios of the accreted material (adjusted for their relative differential sinking) would be very different to Ca/Fe Ni/Fe observed for G200-40, or in fact any other stars. Planetary processes that alter these ratios would then need to be invoked, and it is not clear what these would be.

\subsection{A true match?}
The analysis presented here points towards the many difficulties in comparing abundances derived for cool main-sequence stars to those derived from atmospheric modelling of white dwarfs. Studies of wide binaries that compare abundances between companions can derive very precise differential abundances, particularly if the components of the binaries are very similar to each other \citep[][]{ramirez, tucci_maia2019, Hawkins2020}. Here, our stars are so different from each other that we are limited by various large systematic uncertainties, notably on the white dwarf abundances \citep{Xu2017} and the K-dwarf abundances \citep[see extensive discussions in \eg][]{jofre15}. The uncertainties on the white dwarf abundances result from differences in the derived abundances depending on the stellar parameters (as shown in Table 4 and 5 of \citep{Klein2011}) and abundances that differ depending on the lines used, potentially probing the atmospheric structure of the white dwarf (see \eg Table 3 and 4 of \citep{Jura2012}). We note here that the statistical analysis in the composition comparison was performed assuming random and not systematic errors.

The difference in elemental abundances (Z/Fe) between G200-40 and WD 1425+540 is within $1\sigma$ for all elements, except C/Fe. This fits with a model in which volatiles are depleted in planetary bodies, and indicates a strong match. However, we note here that the abundances for WD 1425+540 would be consistent with many stars selected at random. Our analysis finds that the hypothesis that G200-40 and its companion WD 1425+540 share the same true abundances is more likely than the hypothesis that WD 1425+540 shares the true abundances of a random star analysed in a similar manner in this work, 61 Cyg B. For the former $>70\%$ of randomly selected samples are further apart than the observed abundances, in contrast to $<25\%$ for 61 Cyg B. This conclusion holds if we exclude O (or if we exclude Mg), where the measurement for 61 Cyg B (G200-40) find abundances which are less consistent with other stars of similar metallicity.  However, considering those elements where we are most confident of abundance determinations (Ca, Fe, Si and Ni only), 55\% (70\%) of the random draws are further apart than the observed abundances, given the hypothesis that the true abundances of G200-40 (61 Cyg B) match those of WD 1425+540. In other words, a firm conclusion cannot be reached given the large uncertainties on the observed abundances and further data are required.



\subsection{Implications for exoplanet composition}
If the refractory abundances of an observed exoplanet are known to match those of its host-star, the stellar abundances can be used to improve our knowledge of the planet's composition \citep{Dorn2015}, often constraining in particular the planet's core size \citep{Dorn2017b}. The analysis presented here provides observational evidence in support of this assumption and highlights how stellar abundances can inform our knowledge and understanding of exoplanets.

\section{Conclusions}
In this paper, we present a novel means to investigate whether planetary bodies form with the same composition as their host-stars. Wide binary pairs are found, in general, to be chemically homogeneous \citep{andrews2019, Hawkins2020}. This means that wide binary companions can be used as a proxy for the composition of planet-host stars. Analysis of polluted white dwarfs provide the bulk composition of the exoplanetary material that they have accreted. We compare the abundances of the K-dwarf, G200-40, to the planetary material accreted by its white dwarf companion, WD 1425+540, as a means to compare the composition of rocky exoplanetary material and their host-stars. Elemental abundances (Ca, Ni, Fe, Mg, Si, C and O) of G200-40 are consistent with those of the exoplanetary bodies accreted by its white dwarf companion, WD 1425+540, within the observational errors ($1-\sigma$). In fact, given the observational errors, our analysis finds that in 7/10 observations, the derived abundances for the two objects would on average be further apart than in the observations presented here, even in the case that both the K-dwarf and planetary bodies accreted by its white dwarf companion have the same true abundances. This is in stark contrast to the same analysis for 61 Cyg B, a similar K-dwarf also analysed in this work, where the null hypothesis that the abundances of 61 Cyg B and WD 1425+540 match can be ruled out at a 75\% confidence level. We consider this to be evidence in favour of the hypothesis that exoplanetary bodies have the same refractory composition as their host stars, although noting that our conclusions are limited by the large uncertainties on the data. Our work supports the idea that host-star abundances can be used to improve the determination of the interior of observed rocky exoplanets.

\section{Acknowledgements}
AB acknowledges the support of a Royal Society Dorothy Hodgkin Fellowship. LR is grateful to STFC and the Institute of Astronomy, University of Cambridge for funding her PhD studentship. We thank Marcelo Tucci Maia for preparing the spectrum of G200 for analysis, as well as Laia Casamiquela and Francisca Rojas Espinoza for fruitful discussions. PJ acknowledges partial support of FONDECYT Iniciaci\'{o}n 
grant Number 11170174 and ECOS-ANID collaboration grant  Number 180049. SX acknowledges the supported from the international Gemini Observatory, a program of NSF’s NOIRLab, which is managed by the Association of Universities for Research in Astronomy (AURA) under a cooperative agreement with the National Science Foundation, on behalf of the Gemini partnership of Argentina, Brazil, Canada, Chile, the Republic of Korea, and the United States of America. C.M.\ acknowledges support from the US National Science Foundation grant 
SPG-1826583. The authors wish to thank Bradford Holden for assistance in scheduling the APF observations presented in this paper. Research at Lick Observatory is partially supported by a generous gift from Google.

\subsection{Data Availability Statement}
The data underlying this article will be shared on reasonable request to the corresponding author.

\bibliographystyle{mn}

\bibliography{ref}


\bsp	
\label{lastpage}
\end{CJK}
\end{document}